# Wafer-Scale Epitaxial Modulation of Quantum Dot Density

N. Bart[1]*, C. Dangel[2,3]*, P. Zajac[1], N. Spitzer[1], J. Ritzmann[1], M. Schmidt[1], H. G. Babin[1], R. Schott[1], S. R. Valentin[1], S. Scholz[1], Y. Wang[4], R. Uppu[4], D. Najer[5], M. C. Löbl[5], N. Tomm[5], A. Javadi[5], N. O. Antoniadis[5], L. Midolo[4], K. Müller[3,6], R. J. Warburton[5], P. Lodahl[4], A. D. Wieck[1], J.J. Finley[2,3], and A. Ludwig[1]†

*1- Ruhr-Universität Bochum, Lehrstuhl für Angewandte Festkörperphysik, Universitätsstraße 150, 44801 Bochum, Germany*

*2 – Walter Schottky Institut and Physik Department, Technische Universität München, Am Coulombwall 4, 85748 Garching, Germany*

*3 - Munich Center for Quantum Science and Technology (MCQST), Schellingstr. 4, 80799 Munich, Germany*

*4 - Center for Hybrid Quantum Networks (Hy-Q), Niels Bohr Institute, University of Copenhagen, Blegdamsvej 17, DK-2100 Copenhagen, Denmark*

*5 - Department of Physics, University of Basel, Klingelbergstrasse 82, CH-4056 Basel, Switzerland*

*6 - Walter Schottky Institut and Department of Electrical and Computer Engineering, Technische Universität München, Am Coulombwall 4, 85748 Garching, Germany*

*These authors contributed equally to this work.

†Correspondence to: Arne.Ludwig@rub.de

**Precise control of the properties of semiconductor quantum dots (QDs) is vital for creating novel devices for quantum photonics and advanced opto-electronics. Suitable low QD-density for single QD devices and experiments are challenging to control during epitaxy and are typically found only in limited regions of the wafer. Here, we demonstrate how conventional molecular beam epitaxy (MBE) can be used to modulate the density of optically active QDs in one- and two- dimensional patterns, while still retaining excellent quality. We find that material thickness gradients during layer-by-layer growth result in surface roughness modulations across the whole wafer. Growth on such templates strongly influences the QD nucleation probability. We obtain density modulations between 1 and 10 QDs/µm$^2$ and periods ranging from several millimeters down to at least a few hundred microns. This novel method is universal and expected to be applicable to a wide variety of different semiconductor material systems. We apply the method to enable growth of ultra-low noise QDs across an entire 3-inch semiconductor wafer.**

**Introduction**

Spontaneous pattern formation is common in many natural systems having characteristic sizes ranging from the atomic to the cosmic scale. Typically, spontaneous ordering arises in inherently nonlinear systems due to the complex interplay of thermodynamic and dissipative processes that lead to minimization of local free energies[1]. In the context of the lattice-mismatched growth of III-V semiconductor nanostructures, this principle is exploited to create defect-free nanoscale islands of low bandgap materials surrounded by a wider bandgap matrix, called self-assembled quantum dots (QDs)[2,3]. Such nanostructures are versatile building blocks that are widely used in advanced opto-electronic device technologies, such as highly performant LEDs[4] and energy efficient nano-lasers[5], as well as discrete quantum components like non-classical light sources for use in photonic quantum technologies[6-8]. Key factors for the device integration of such QDs are their size, shape and composition, and control of their areal density[9]. For example, exploiting their narrow emission linewidth for modal gain in nano-lasers requires high density regions to provide sufficient gain[10], whereas in quantum technology, highly-efficient single-photon sources require low density and positioning over the length scale of the optical wavelength[11,12]. Key metrics for the QD quality are near-transform limited emission and absorption linewidths and near-unity single photon indistinguishability (see Ref [6] and references therein).

The mechanism that drives self-assembled QD growth is based on strain relaxation during heteroepitaxy of materials having different lattice constants. In the case of InAs on GaAs, strain builds up due to the 7% larger lattice constant of InAs compared to GaAs, inducing a change of growth mode from layer-by-layer growth (Frank-Van der Merwe) to layer-plus-island Stranski-Krastanov (SK) growth[13,14]. The exact moment of nucleation is heavily influenced by the growth conditions[15,16]. Due to a steep onset of the nucleation at a critical InAs layer thickness[14], low QD density control is challenging.

We find that controlling the surface roughness at the atomic scale is a key factor for engineered QD nucleation that has been largely neglected until now. Compared to atomically smooth growth surfaces, rougher surfaces enhance the QD nucleation probability[17-19]. It is well known that atomically flat substrates successively undergo cycles of roughening and smoothening as the fractional completion of each monolayer changes; non-integer filling of each atomic layer results in atomically rough surfaces, that smoothen as the monolayer is completed. This property is utilized, for example, in reflection high-energy electron diffraction (RHEED) growth rate analysis, a standard method in e.g. MBE[20].

Here, we exploit the impact of roughness on QD-nucleation by growing layer thickness gradients prior to the deposition of QDs, thus creating *in-situ* integer / non-integer layer numbers and roughness modulations. As a result, QD density modulations over the entire wafer are created. By controlling (i) the orientation of the substrate relative to the effusion cell, (ii) the deposition amount and interrupt time and (iii) the substrate temperature, we show that we can precisely engineer the roughness distribution to produce a variety of QD density patterns on the wafer in one and two dimensions.

**Results**

The key step in our sample preparation is the growth of a gradient layer, that is termed a pattern defining layer (PDL). This is illustrated in Fig. 1a: by depositing material from an inclined effusion cell while substrate rotation is stopped, a thickness gradient is created. We grow such a PDL consisting of a GaAs gradient layer with a nominal thickness of 15 nm at the wafer center, corresponding to an overall thickness difference of 22 monolayers (ML) across the entire wafer. After deposition of the PDL, the substrate temperature was reduced from 600 °C to 525 °C, thereby (i) enabling InAs deposition without excessive desorption[21] and (ii) preserving the surface morphology of the PDL. On top of the GaAs PDL, self-assembled QDs were grown by depositing InAs and subsequently capped with GaAs (refer to Ludwig et al.[22] for QD growth details and method section for further preparation).

A typical ensemble photoluminescence (PL) spectrum recorded from such QDs is presented in Fig. 1b. It exhibits three distinct emission peaks corresponding to parity allowed interband transitions between the orbital states in the QDs. Fig. 1c shows a typical map of the PL intensity of the entire QD emission integrated within the spectral range of 1000 nm – 1300 nm at each point of a wafer. The data reveal a clear modulation of the integrated QD emission intensity along the x-direction, seen in Fig 1c as a curved stripe pattern. To investigate the impact of the surface roughness on the local QD density, we performed a series of annealing tests in which the substrate and PDL was held for a time $t_{anneal}$ at $T_{substrate} = 600$ °C in order to smooth it before deposition of the QD layer. Figure 1c compares data recorded for different annealing times immediately before the substrate temperature was reduced for the InAs deposition. In contrast to the clear modulation for the $t_{anneal} = 0$ s reference sample, the pattern progressively disappears after $t_{anneal} = 210$ s of annealing. For the longest investigated annealing time of $t_{anneal} = 600$ s, all intensity modulation other than the inhomogeneity due to the inherent indium cell flux distribution disappears. We observe the strongest intensity modulation in regions of just the critical amount of deposited InAs for QD nucleation and, for this coverage the local QD density varies from zero to finite values.

To quantitatively compare the patterns, we calculated the Michelson contrast at comparable densities for the different samples (Supp. Information). Plotting this contrast over the annealing time in Fig. 1d, we observe a stable contrast for the first 60 s, after which it diminishes. The complete disappearance of the QD density modulation for an annealing time of 600 s is comparable to the coalescence time of surface islands and holes reported by Franke et al.[23] in surface morphology studies. This observation suggests a link between the local surface roughness on the wafer and the QD density, similar to studies of growth on vicinal substrates[24,25].

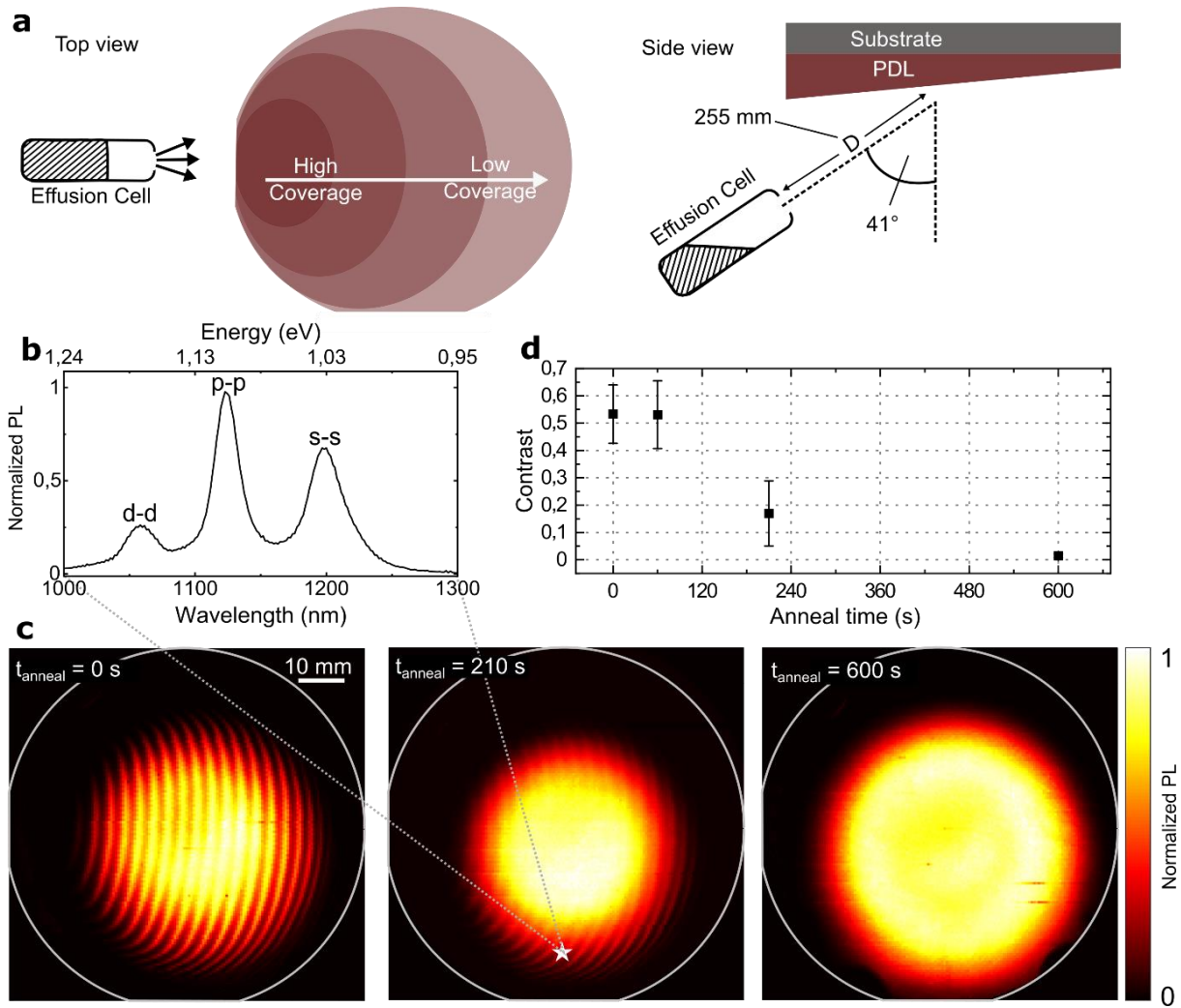

**Fig. 1 | Effusion cell geometry and QD density modulation. a**, Schematic representation of the gradient of material coverage on the substrate in top view (left) and geometrical configuration of the Ga-effusion cell inside the MBE growth chamber viewed from the side (right). **b**, Ensemble photoluminescence (PL) spectrum at 77 K with a laser spot size of ~ 100 µm from a 210 s annealed sample (white star). The different peaks correspond to the different dipole and parity allowed interband transitions between orbital states. **c**, False color PL maps recorded from 3" wafers with a nominally 15 nm thick GaAs pattern defining layer (PDL). The QD PL intensity is spectrally integrated over the region between 1000 - 1300 nm. The wafers were annealed before the QD growth for 0 s, 210 s and 600 s, respectively. **d**, Michelson contrast at medium densities over the annealing time.

To confirm this expectation, we performed experiments in which growth was stopped before deposition of the QD layer, thereby creating a GaAs PDL on the surface. Atomic force microscopy (AFM) measurements were performed on the surface at multiple points along the thickness gradient. Figure 2a shows typical data recorded at different stages of GaAs coverage relative to a smooth (integer value) surface. We define an integer number of layers, by choosing an arbitrary starting point of 0 ML at the location where the lowest step density is measured and observe a clear progression of the surface in accordance with layer-by-layer growth, similar to surface studies by Bell et al.[26]. Completely finished monolayers, termed 0 ML here, show smooth surfaces covered with only a few small islands and holes. The wide monolayer terraces as seen in the first image (0 ML) occur due to the unintentional wafer miscut and consist of monolayer steps. The width of the steps is on average 500 nm, which corresponds to a 0.03° miscut. Increasing the coverage by 0.25 ML results in a formation of ~ 60 nm wide islands which are elongated along the [0$\bar{1}$1] direction. At 0.5 ML coverage, these islands merge which makes the wafer miscut only barely visible. This finding indicates that step flow growth, meaning the growth of the miscut related preexisting steps, is insignificant. Adatoms do not accumulate at the miscut steps but rather nucleate on top of them. Lastly, at around 0.75 ML the merged islands leave behind small, elongated gaps. The step densities along the [011] direction determined from these AFM images are presented in Fig. 2b, exhibiting a variation between 6 and 13 steps/µm. The lowest measured step density is still higher than the ~1.4 steps/µm in horizontal direction stemming from the miscut, since some finite roughness (islands or holes) always exists if the surface is not annealed. Comparing the density modulation periodicity of 3 mm observed in these AFM measurements with the periodicity of the PL intensity, it is clear that each stripe in the PL maps is the result of the variation in step density between two integer monolayers of GaAs of the underlying PDL.

To prove that the PL intensity pattern is a direct consequence of the modulated QD density and to examine morphology of such QDs, we performed AFM measurements of another sample, where the growth was stopped immediately after the deposition of the InAs QD layer.

Fig. 3a shows typical AFM measurements performed on these samples along the thickness gradient in a region of low QD density. For the underlying In(Ga)As wetting layer, we observe a similar modulation of the atomic island roughness arrangement as for the GaAs PDL of the sample discussed in relation to Fig. 2. In contrast to the GaAs surface, the In(Ga)As shows much larger islands which results from the increased surface diffusion length of InAs[14].

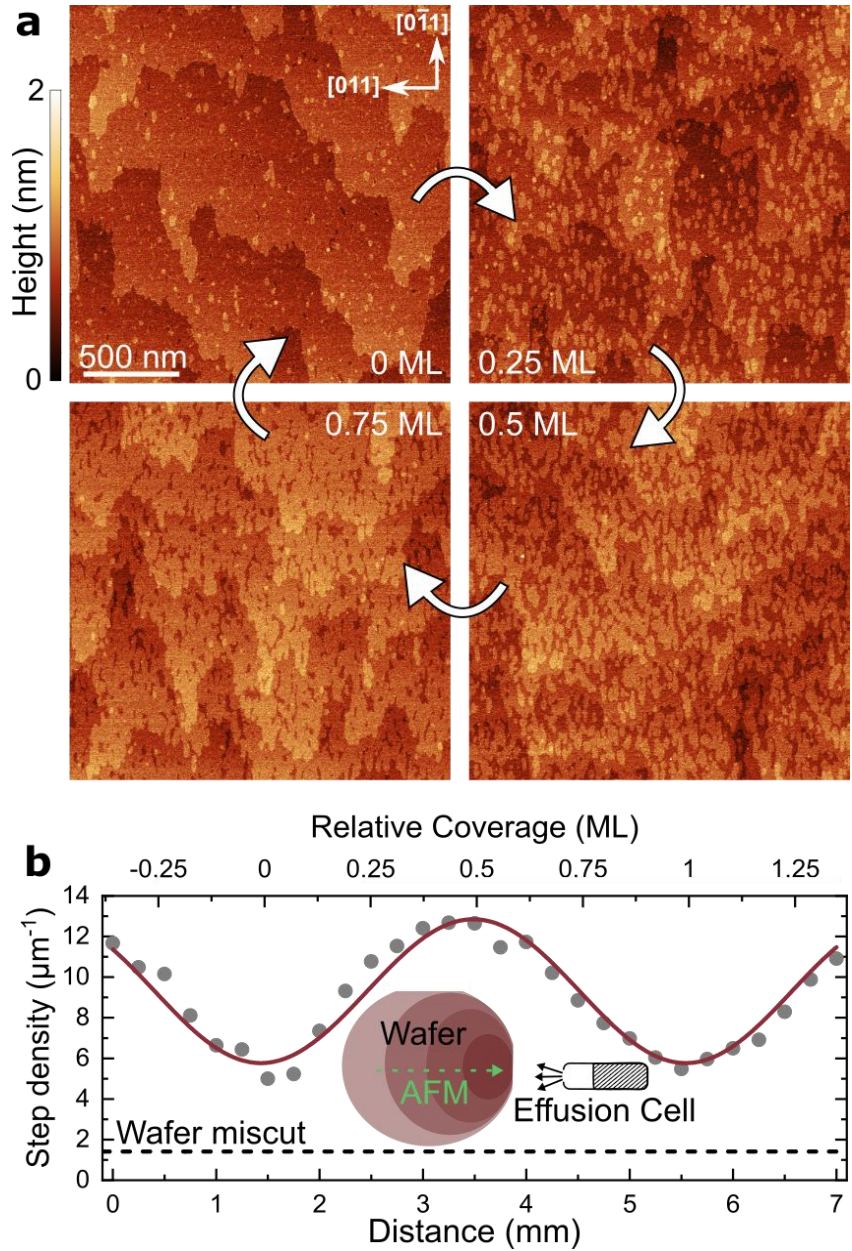

**Fig. 2 | Atomic force microscopy measurements of a surface pattern defining layer. a**, AFM maps of GaAs surfaces after a coverage of 0, 0.25, 0.5 and 0.75 ML relative to the location of lowest step density. **b**, Step density (circles) determined from AFM measurements along the thickness gradient of the PDL (green dashed line in the inset wafer illustration) and sinusoidal fit (red line). The green circles correspond to the four images in **a**. The calculated step density originating from the wafer miscut in horizontal [011] direction is marked by the dashed black line and is determined from average terrace widths.

However, the most striking observation in these data pertains to the modulation of the QD density. The deposited InAs is barely enough to induce QD nucleation (~ 1.6 ML), as evident from the overall low QD density and the presence of a second species of smaller QDs[27]. The density of the larger QDs, to which we attribute the observed PL signal, is plotted in Fig. 3b and is modulated between ~ 1 and 10 QDs/µm². We find that the QDs tend to be slightly larger in size at low density regions (Supp. Information). Furthermore, we do not observe a step erosion of the wetting layer as described by Placidi et al.[28] or preferred QD nucleation at the step edges as described by Leon et al.[25], since most QDs seem to be on terraces away from step edges. This observation implies that the dominant process for nucleation can be traced to filling of holes in the GaAs PDL by InAs.

As illustrated schematically in Fig. 3c, we attribute the increased QD density to a local reduction in the effective critical InAs amount for QD nucleation. When depositing InAs on top of GaAs surfaces, layer-by-layer growth occurs on smooth surfaces, while the holes observed in the AFM images are filled by InAs and subsequently overgrown by extended monolayers of InAs. As a result of the effective increase in the local layer thickness, growth at the InAs filled locations experience a higher strain than the thinner layers on GaAs where holes are not present. Thus, the strain induced QD nucleation is more likely to occur above such holes. As a consequence, hole-dominated surfaces, i.e. with GaAs surface coverage > 0.5 ML tend to show higher QD density, while smooth or island dominated surfaces with a coverage < 0.5 ML GaAs show smaller QD densities, resulting in a modulation of the QD density. The step density of the underlying GaAs surface is hidden by the In(Ga)As wetting layer. Hence, determining the precise roughness of the PDL from the wetting layer is challenging due to the difficulty of determining (i) intermixing of In and Ga in the wetting layer, (ii) re-evaporation of In and (iii) the exact InAs amount used for QD nucleation.

We continue to demonstrate control over the pattern formation by tailoring nominal thickness gradients along different axes of the 3-inch wafer. Fig. 4a shows that doubling the PDL thickness from 15 nm to 30 nm halves the modulation period, from 3 mm to 1.5 mm. This provides a unique way to directly measure the PDL cell effusion profile of any MBE system with sub-monolayer precision across the entire wafer, simply by recording the spatial dependence of the QD PL intensity.

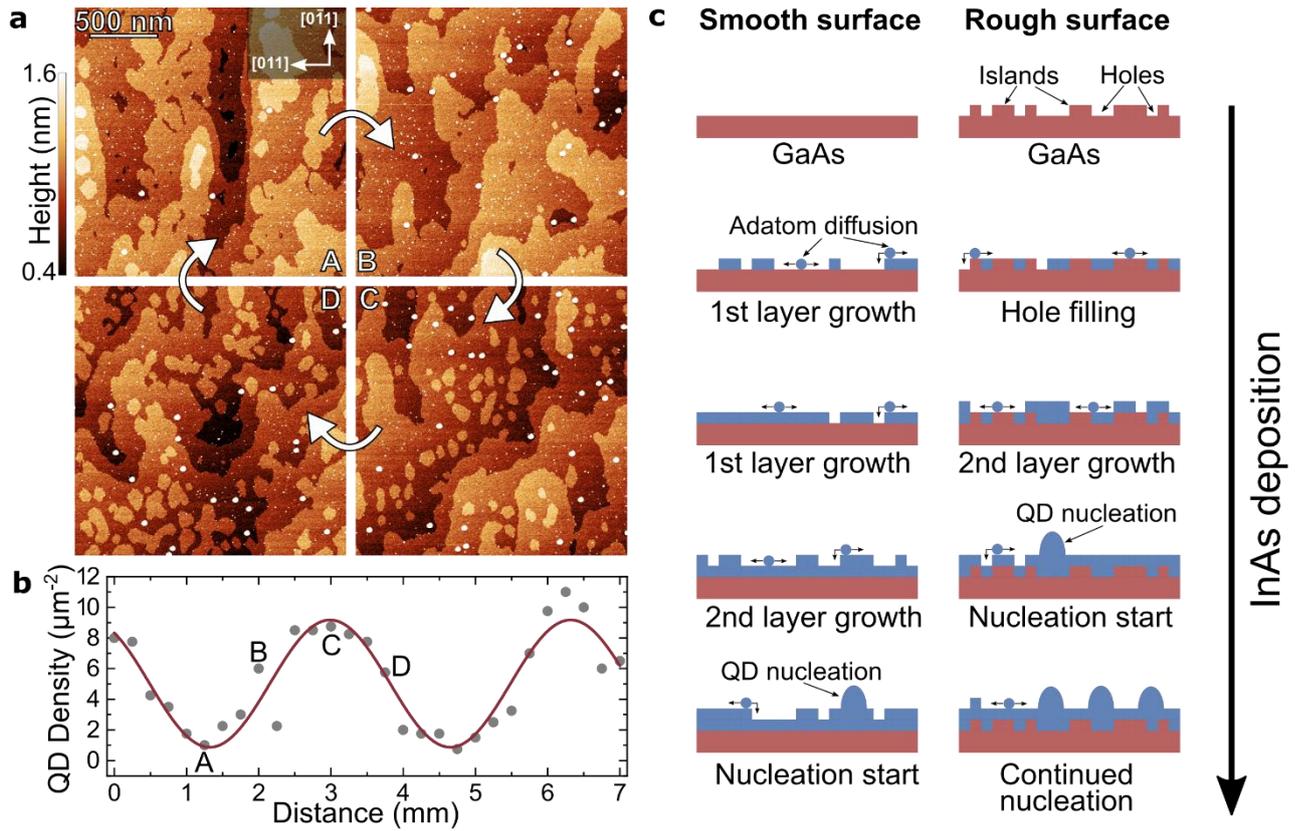

**Fig. 3 | Atomic force microscopy measurements of surface QDs and enhanced nucleation schematic. a**, AFM images of surface QDs along the PDL GaAs gradient direction. **b**, QD densities determined from AFM images along the PDL direction (grey dots) and sinusoidal fit (red line). The images in **a** are marked by the corresponding letters and represent one PDL, i.e. GaAs ML cycle. **c,** Schematic illustrating InAs layer (blue) development under increasing InAs deposition on a smooth and rough GaAs surface (red). Adatom diffusion (blue dots) takes place on the surface. QD nucleation (blue domes) on rough surfaces starts earlier than on smooth surfaces.

The data presented in Fig. 4b was recorded from a sample for which we first grew an 80 nm thick PDL to define a specific axis along which the density is modulated, followed by a 60 s smoothing growth interruption, before growing a second 40 nm PDL along an axis orientated at a relative angle of 120° to the first. The smoothing growth interruption between the growth of the two PDLs is necessary to provide partially smoothed areas interspersed in rough regions for the second layer modulation while still retaining some of the roughness modulation of the first layer. These observations clearly show that our methods are highly flexible, allowing design of a specific 2D pattern across the entire wafer before QD growth.

We now continue to demonstrate that our method is equally applicable to PDLs formed in ternary alloys. Hereby, we present in Fig. 4c QD PL data recorded from a PDL defined by depositing 150 nm of $Al_{0.33}Ga_{0.67}As$ and a 2.5 nm thick GaAs buffer layer before depositing the QDs. Clearly the underlying PDL roughness modulation is preserved. Similar results were obtained using a pure AlAs PDL (Supp. Information). Furthermore, the data presented in Fig. 4c shows how, by increasing the PDL thickness from 30 nm to 150 nm, the modulation period can be further reduced to 300 µm. This demonstrates that the roughness modulation is preserved at the growth surface, even after >500 MLs have been deposited. For instance, this is much more than would be observable in RHEED oscillations for the specific growth conditions used here.

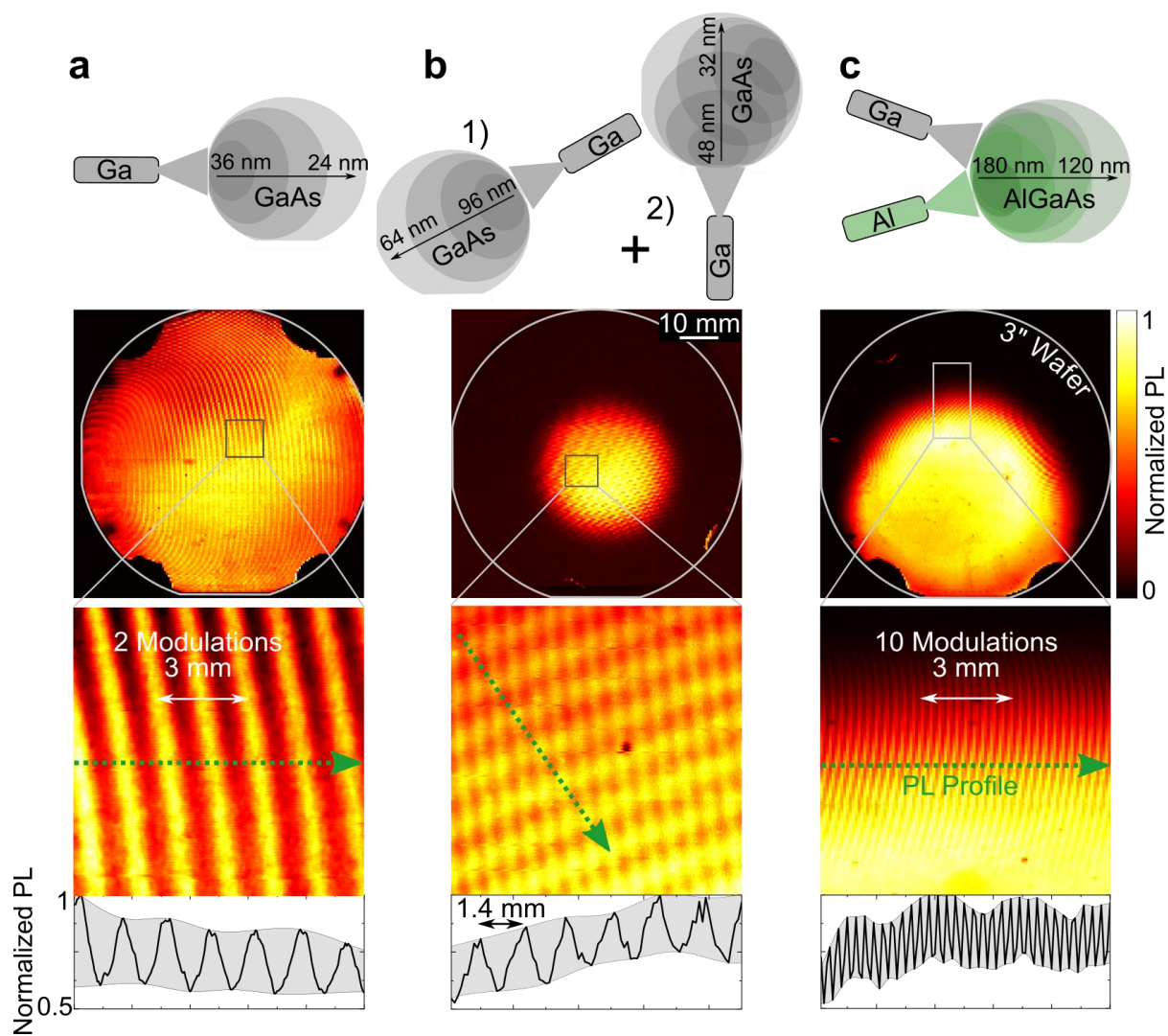

**Fig. 4 | Demonstration of epitaxial pattern control.** QD PL intensity maps of **a,** 30 nm GaAs PDL, **b,** superposition of 80 and 40 nm GaAs PDL, and **c,** 150 nm AlGaAs PDL. White/yellow indicate high intensity, red/black low. High resolution maps of the marked areas are shown below. The normalized PL intensity along the green dotted line of the respective zoom-ins is presented in the bottom row.

The data presented in Fig. 4 clearly shows that spatial roughness modulation is an exceptionally useful tool to achieve low QD densities on a full wafer. Furthermore, the approach is universal and can be used for gradient layer-by-layer growth using any binary or alloy in the group-III arsenide family. Thus, we conclude that the roughness modulation method presented here should also be fully applicable to other materials systems that involve strain-driven self-assembly[13]. Beyond this, first growth trials using our MBE system hint towards ordering effects of metallic droplets on an AlGaAs PDL which could, for example, be used for droplet epitaxy or hole etching and subsequent local droplet-etched dot growth[29-31] (Supp. Information).

Even smaller modulation periods than those presented in Fig. 4 could be achieved by using steeper material gradients. This would require either thicker layers, shallower angles of the wafer relative to the cell, or partial flux shadowing. We anticipate that the physical limitation of our method is most likely defined by the point at which the Ga adatom diffusion length on GaAs ($< 10$ nm[32]) becomes comparable to the roughness modulation period. This makes us confident that epitaxial control of QD nucleation at sub-micron length scales is reachable, possibly down to the optical wavelength in the medium.

Achieving *in-situ* alignment on this length scale has the potential to be transformative for quantum technology applications, since QDs grown using the scheme presented here have shown near-transform limited linewidths and near-unity photon indistinguishability proving the excellent optical quality (Supp. Information and Refs[11,12,33-41]). We believe this low-noise environment is facilitated by the absence of prolonged growth breaks by preventing the incorporation of impurities and creation of crystal defects[42] that can thermally trap and release charge carriers.

We note that it is likely that our PDL technique is not compatible with step-flow growth since no roughness variations occur. However, we believe the roughness modulation method presented in this work clearly demonstrates a simple and efficient way to control the density of low-noise, high-quality self-assembled QD nanostructures for advanced opto-electronic and quantum photonic applications.

# Acknowledgements

The authors thank Johanna Eichhorn for help with AFM measurements. Moreover, we gratefully acknowledge support from the German Federal Ministry of Education and Research via the funding program Photonics Research Germany (contract number 13N14846, QR.X Project 16KISQ009 and 16KISQ027). Furthermore, C.D., K.M, J.J.F. gratefully acknowledge the Deutsche Forschungsgemeinschaft (DFG, German Research Foundation) via Germany's Excellence Strategy, via e-conversion EXC-2111/1 – 390814868 and MCQST EXC-2089/1 – 390776260. J.J.F. gratefully acknowledges the DFG for funding via projects FI947-5, FI947-6 and INST 95-164. N.B., A.D.W., and A.L. acknowledge gratefully support of TRR 160/2-Project B04, DFG 383065199 and the DFH/UFA CDFA-05-06.

# Competing interests

N.B., C.D., K.M., A.D.W., J.J.F., and A.L. applied for European patent under file number EP19177713.

# Methods

## Sample growth

All samples were grown on undoped (100) surfaces of 3" GaAs wafers with a miscut <0.1 ° (as specified by the vendor) using a custom horizontal MBE System. Before growth, wafers were heated to 640 °C under an arsenic atmosphere of $9.6 \cdot 10^{-6}$ Torr beam equivalent pressure (BEP) to remove surface oxides. An arsenic valved cracker was employed, operating at 700 °C, providing primarily As$_4$. We used growth rates of 0.2 nm/s for GaAs, 0.1 nm/s for AlAs and ~0.013 nm/s for InAs. We prepared the wafers by deposition of a buffer consisting of a 50 nm thick GaAs layer and a 30 period superlattice of 2 nm AlAs and 2 nm GaAs, followed by another 50 nm GaAs buffer layer, all grown at 600 °C and an As BEP of $9.6 \cdot 10^{-6}$ Torr. For electrical contact, we grew a Si-doped back contact with a doping concentration of $2 \cdot 10^{18}$ cm$^{-3}$, followed by a 5 min annealing break and a 5 nm GaAs layer at 575 °C to prevent silicon segregation. After an increase back to 600 °C, a pattern defining layer (PDL) of GaAs, AlAs or a ternary alloy $Al_xGa_{(1-x)}As$ was grown (Supp. Information Table 1). To avoid direct QD growth on ternary alloys, a 2.5 nm thick spacer layer of GaAs was deposited on the Al-containing PDLs. After the PDL deposition, the substrate temperature is reduced from 600 °C to 525 °C by 50 °C in 30 s, another 25 °C in 60 s and followed by a 60 s settling break. Quantum dots are grown in SK-growth mode at 525 °C substrate temperature and an As BEP of $6.8 \cdot 10^{-6}$ Torr. For this, InAs is deposited in cycles of 4 s growth, followed by a 4 s break, amounting to a total of 12 cycles and resulting in coverages of 1.6 – 1.8 ML. During the first 2 – 4 cycles, the wafer rotation is stopped so that the indium effusion cell is oriented towards the wafer big flat. After an additional 20 s break, the QDs are then capped with a 10 nm thick layer of undoped GaAs, 130 nm AlGaAs and 5 nm GaAs. QDs in samples where the dot height was reduced to 3 nm due to the indium flushing method have an emission wavelength of 910 to 960 nm. For more sample and QD growth details, the reader is referred to Ludwig et al.[22].

## Photoluminescence measurements

### Wafer mapping

Photoluminescence (PL) measurements were performed by exciting samples with a 518 nm laser with a spot size of ~ 100 µm in diameter with total excitation powers between 1 and 20 mW. Liquid nitrogen was used to cool a 3" cold-finger inside a cryostat which is fixed to two stepping motors for position control. Thus, sample temperature for all PL measurements is approximately 100 K. A spectrometer equipped with a Si-CCD was used for measurement of wavelengths between 340 and 1020 nm, combined with an InGaAs line array detector for 900 to 1715 nm.

## Atomic force microscopy

For atomic force microscopy (AFM) measurements, a Bruker Dimension Icon system was used in PeakForce tapping mode. Areas of 2x2 µm² with a resolution of 512x512 $px^2$ were scanned.

The step density is extracted by counting how many times the derivative along the [011] direction surpasses a set threshold for each measured line during AFM.

# References

1. Teichert, C. Self-organization of nanostructures in semiconductor heteroepitaxy. *Physics Reports* **365**, 335-432 (2002).
2. Marzin, J.-Y., Gérard, J.-M., Izraël, A., Barrier, D. & Bastard, G. Photoluminescence of single InAs quantum dots obtained by self-organized growth on GaAs. *Physical review letters* **73**, 716 (1994).
3. Moison, J. *et al.* Self-organized growth of regular nanometer-scale InAs dots on GaAs. *Applied Physics Letters* **64**, 196-198 (1994).
4. Liu, Z. *et al.* Micro-light-emitting diodes with quantum dots in display technology. *Light Sci Appl* **9**, 83, doi:10.1038/s41377-020-0268-1 (2020).
5. Strauf, S. & Jahnke, F. Single quantum dot nanolaser. *Laser & Photonics Reviews* **5**, 607-633, doi:10.1002/lpor.201000039 (2011).
6. Senellart, P., Solomon, G. & White, A. High-performance semiconductor quantum-dot single-photon sources. *Nat Nanotechnol* **12**, 1026-1039, doi:10.1038/nnano.2017.218 (2017).
7. Lodahl, P. Quantum-dot based photonic quantum networks. *Quantum Science and Technology* **3**, 013001 (2017).
8. Huber, D., Reindl, M., Aberl, J., Rastelli, A. & Trotta, R. Semiconductor quantum dots as an ideal source of polarization-entangled photon pairs on-demand: a review. *Journal of Optics* **20**, 073002 (2018).
9. Grydlik, M., Langer, G., Fromherz, T., Schaffler, F. & Brehm, M. Recipes for the fabrication of strictly ordered Ge islands on pit-patterned Si(001) substrates. *Nanotechnology* **24**, 105601, doi:10.1088/0957-4484/24/10/105601 (2013).
10. Kreinberg, S. *et al.* Emission from quantum-dot high-beta microcavities: transition from spontaneous emission to lasing and the effects of superradiant emitter coupling. *Light Sci Appl* **6**, e17030, doi:10.1038/lsa.2017.30 (2017).
11. Uppu, R. *et al.* Scalable integrated single-photon source. *Science Advances* **6**, eabc8268 (2020).
12. Tomm, N. *et al.* A bright and fast source of coherent single photons. *Nature Nanotechnology*, 1-5 (2021).
13. Sautter, K. E., Vallejo, K. D. & Simmonds, P. J. Strain-driven quantum dot self-assembly by molecular beam epitaxy. *Journal of Applied Physics* **128**, 031101 (2020).
14. Leonard, D., Pond, K. & Petroff, P. M. Critical layer thickness for self-assembled InAs islands on GaAs. *Phys Rev B Condens Matter* **50**, 11687-11692, doi:10.1103/physrevb.50.11687 (1994).
15. Vannarat, S., Sluiter, M. H. & Kawazoe, Y. Effect of strain on alloying in InAs/GaAs heterostructure. *Japanese journal of applied physics* **41**, 2536 (2002).
16. Heyn, C. Critical coverage for strain-induced formation of InAs quantum dots. *Physical Review B* **64**, 165306 (2001).
17. Xu, M. C., Temko, Y., Suzuki, T. & Jacobi, K. On the location of InAs quantum dots on GaAs(001). *Surface Science* **589**, 91-97, doi:10.1016/j.susc.2005.05.052 (2005).
18. Wang, Z. M. *et al.* Localized formation of InAs quantum dots on shallow-patterned GaAs(100). *Applied Physics Letters* **88**, doi:10.1063/1.2209157 (2006).

19　　Borgström, M., Johansson, J., Landin, L. & Seifert, W. Effects of substrate doping and surface roughness on self-assembling InAs/InP quantum dots. *Applied surface science* **165**, 241-247 (2000).

20　　Hermann, M. & Sitter, H. Molecular beam epitaxy. *Springer* (1996).

21　　Heyn, C. Stability of InAs quantum dots. *Physical Review B* **66**, 075307 (2002).

22　　Ludwig, A. *et al.* Ultra-low charge and spin noise in self-assembled quantum dots. *Journal of Crystal Growth* **477**, 193-196 (2017).

23　　Franke, T., Kreutzer, P., Zacher, T., Naumann, W. & Anton, R. In situ RHEED, AFM, and REM investigations of the surface recovery of MBE-grown GaAs (0 0 1)-layers during growth interruptions. *Journal of crystal growth* **193**, 451-459 (1998).

24　　Schmidbauer, M. *et al.* Controlling planar and vertical ordering in three-dimensional (In,Ga)As quantum dot lattices by GaAs surface orientation. *Phys Rev Lett* **96**, 066108, doi:10.1103/PhysRevLett.96.066108 (2006).

25　　Leon, R., Senden, T., Kim, Y., Jagadish, C. & Clark, A. Nucleation transitions for InGaAs islands on vicinal (100) GaAs. *Physical review letters* **78**, 4942 (1997).

26　　Bell, G., Jones, T., Neave, J. & Joyce, B. Quantitative comparison of surface morphology and reflection high-energy electron diffraction intensity for epitaxial growth on GaAs. *Surface science* **458**, 247-256 (2000).

27　　Arciprete, F. *et al.* How kinetics drives the two-to three-dimensional transition in semiconductor strained heterostructures: The case of In As∕Ga As (001). *Applied physics letters* **89**, 041904 (2006).

28　　Placidi, E. *et al.* Step erosion during nucleation of InAs∕GaAs (001) quantum dots. *Applied Physics Letters* **86**, 241913 (2005).

29　　Gurioli, M., Wang, Z., Rastelli, A., Kuroda, T. & Sanguinetti, S. Droplet epitaxy of semiconductor nanostructures for quantum photonic devices. *Nature materials*, 1 (2019).

30　　Babin, H. G. *et al.* Charge Tunable GaAs Quantum Dots in a Photonic nip Diode. *Nanomaterials* **11**, 2703 (2021).

31　　Zhai, L. *et al.* Low-noise GaAs quantum dots for quantum photonics. *Nature Communications* **11**, 4745, doi:10.1038/s41467-020-18625-z (2020).

32　　Ohta, K., Kojima, T. & Nakagawa, T. Anisotropic surface migration of Ga atoms on GaAs (001). *Journal of Crystal Growth* **95**, 71-74 (1989).

33　　Pedersen, F. T. *et al.* Near Transform-Limited Quantum Dot Linewidths in a Broadband Photonic Crystal Waveguide. *ACS Photonics* **7**, 2343-2349, doi:10.1021/acsphotonics.0c00758 (2020).

34　　Uppu, R. *et al.* On-chip deterministic operation of quantum dots in dual-mode waveguides for a plug-and-play single-photon source. *Nature Communications* **11**, 3782, doi:10.1038/s41467-020-17603-9 (2020).

35　　Appel, M. H. *et al.* Coherent Spin-Photon Interface with Waveguide Induced Cycling Transitions. *Physical Review Letters* **126**, 013602 (2021).

36　　Najer, D. *et al.* A gated quantum dot strongly coupled to an optical microcavity. *Nature* **575**, 622-627, doi:10.1038/s41586-019-1709-y (2019).

37　　Kurzmann, A., Ludwig, A., Wieck, A. D., Lorke, A. & Geller, M. Auger Recombination in Self-Assembled Quantum Dots: Quenching and Broadening of the

|    | Charged Exciton Transition. *Nano Letters* **16**, 3367-3372, doi:10.1021/acs.nanolett.6b01082 (2016). |
|----|---|
| 38 | Kurzmann, A. *et al.* Optical blocking of electron tunneling into a single self-assembled quantum dot. *Physical review letters* **117**, 017401 (2016). |
| 39 | Kurzmann, A., Ludwig, A., Wieck, A. D., Lorke, A. & Geller, M. Photoelectron generation and capture in the resonance fluorescence of a quantum dot. *Applied Physics Letters* **108**, 263108 (2016). |
| 40 | Kurzmann, A. *et al.* Optical detection of single-electron tunneling into a semiconductor quantum dot. *Physical Review Letters* **122**, 247403 (2019). |
| 41 | Lochner, P. *et al.* Contrast of 83% in reflection measurements on a single quantum dot. *Scientific Reports* **9**, 8817, doi:10.1038/s41598-019-45259-z (2019). |
| 42 | Lan, H. & Ding, Y. Ordering, positioning and uniformity of quantum dot arrays. *Nano Today* **7**, 94-123 (2012). |

# Supplementary Information to:

# Wafer-Scale Epitaxial Modulation of Quantum Dot Density

N. Bart[1]*, C. Dangel[2,3]*, P. Zajac[1], N. Spitzer[1], J. Ritzmann[1], M. Schmidt[1], H. G. Babin[1], R. Schott[1], S. R. Valentin[1], S. Scholz[1], Y. Wang[4], R. Uppu[4], D. Najer[5], M. C. Löbl[5], N. Tomm[5], A. Javadi[5], N. O. Antoniadis[5], L. Midolo[4], K. Müller[3,6], R. J. Warburton[5], P. Lodahl[4], A. D. Wieck[1], J.J. Finley[2,3], and A. Ludwig[1†]

*1- Ruhr-Universität Bochum, Lehrstuhl für Angewandte Festkörperphysik, Universitätsstraße 150, 44801 Bochum, Germany*

*2 – Walter Schottky Institut and Physik Department, Technische Universität München, Am Coulombwall 4, 85748 Garching, Germany*

*3 - Munich Center for Quantum Science and Technology (MCQST), Schellingstr. 4, 80799 Munich, Germany*

*4 - Center for Hybrid Quantum Networks (Hy-Q), Niels Bohr Institute, University of Copenhagen, Blegdamsvej 17, DK-2100 Copenhagen, Denmark*

*5 - Department of Physics, University of Basel, Klingelbergstrasse 82, CH-4056 Basel, Switzerland*

*6 - Walter Schottky Institut and Department of Electrical and Computer Engineering, Technische Universität München, Am Coulombwall 4, 85748 Garching, Germany*

*These authors contributed equally to this work.

†Correspondence to: *Arne.Ludwig@rub.de*

**Supplementary Table 1 | Overview of sample growth parameters. PDL is the pattern defining layer.**

| Figure | PDL | PDL Temperature (°C) | Smoothing break (s) | Wafer size (") | Flushing height (nm) | Internal # |
|---|---|---|---|---|---|---|
| 1c (left) | 15 nm GaAs | 600 | 600 | 3 | - | 15155 |
| 1c (middle) | 15 nm GaAs | 600 | 210 | 3 | - | 15167 |
| 1c (right), S3 | 15 nm GaAs | 600 | 0 | 3 | - | 15154 |
| 2 | 15 nm GaAs | 600 | 0 | 3 | - | 15435 |
| 3, S10 | 15 nm GaAs | 600 | 0 | 3 | - | 15424 |
| 4a | 30 nm GaAs | 600 | 0 | 3 | 3 | 15258 |
| 4b | 150 nm Al33Ga67As + 2.5 nm GaAs | 600 | 0 | 3 | 3 | 15095 |
| 4c | 1) 80 nm GaAs, 2) 40 nm GaAs | 600 | 1) 60, 2) 0 | 3 | - | 15189 |
| S1a, S2 | 15 nm GaAs | 525 | 30 | 3 | 3 | 15097 |
| S1b | 75 nm Al33Ga67As + 2.5 nm | 600 | 0 | 3 | 3 | 15074 |
| S1c | 15 nm AlAs +2.5 nm GaAs | 600 | 0 | 3 | 3 | 15088 |
| S4, S5 | 15 nm GaAs | 600 | 0 | 3 | - | 15288 |
| S6 | 30 nm AlGaAs | 630 | 30 + 120 | 3 | - | 15182 |
| S7 | 31 nm GaAs | 600 | 0 | 3 | 2.4 | 14843 |
| S8 | 35 nm GaAs + 10 nm AlGaAs | 600 | 30 | 3 | 2.8 | 14813 |
| S9 | 40 nm GaAs | 600 | 30 | 3 | 2.2 | 15027 |

**Additional samples with different Al-concentration and geometry**

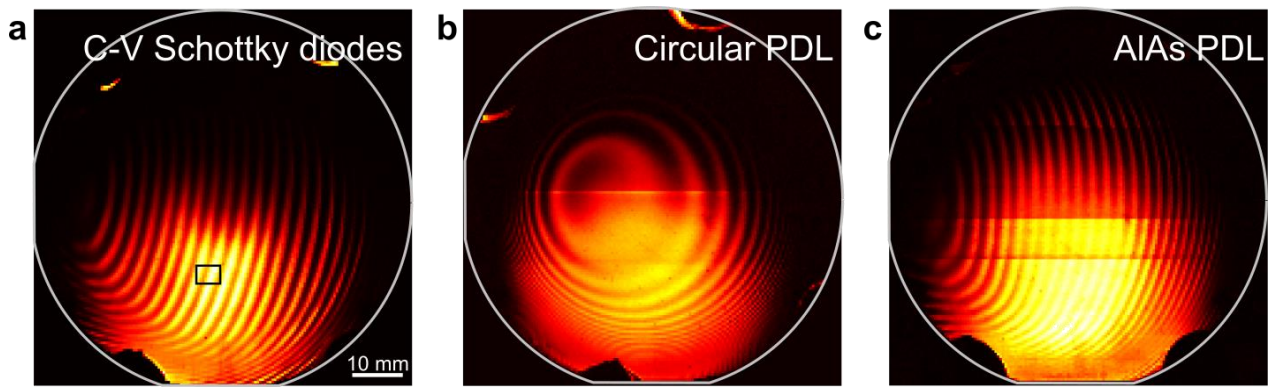

**Supp. Fig. 1 | Photoluminescence (PL) measurements performed on additional samples.** **a**, Sample wafer used for C-V measurement presented in Supp. Fig. 2. The C-V data presented in Supp. Fig. 2 was recorded from Schottky diode samples processed from the marked region on the wafer. **b**, 75 nm thick $Al_{33}Ga_{67}As$ PDL with wafer rotation resulting in a circular pattern. **c**, 15 nm thick AlAs PDL. Horizontal line discontinuities in the PL intensity maps originate from interrupted cooling during the line-scan PL map.

**QD density measurements using capacitance voltage spectroscopy**

To further support that the islands observed in the AFM measurements are the relevant optically active QDs, we performed capacitance-voltage (C-V) spectroscopy measurements. The obtained capacitive signal is directly proportional to the absolute number of QDs below the gate electrode (see below for a description of the samples). In Supp. Fig. 2a we present typical results obtained from devices fabricated at locations on the wafer having the lowest and highest QD densities $\rho_{QD}$, respectively. We observe two charging peaks of a first and a second electron in each QD of a Schottky diode sample. C-V integrated over the voltage range corresponding to the inhomogeneously broadened charging peak increases as an increasing number of QDs can be charged. $\rho_{QD}$ values extracted from this are plotted in Supp. Fig. 2b along with the PL intensity at these locations. In agreement with the PL intensity modulation, the obtained local QD densities at this specific area on the wafer range from 4 µm$^{-2}$ to 11 µm$^{-2}$, similar to the values obtained from uncapped QDs in AFM-measurements (cf. Fig. 3).

For C-V measurements, n-i-Schottky diodes were processed from the wafer piece marked with a black outline in Supp. Fig. 1a. n-contacts are formed by indium-solder to the sample corners. The Schottky contacts are 100 nm thick 300 x 300 µm² gold gates, with a 2 nm chrome adhesion layer, that are bonded inside a 16 pin carrier, using ultrasonic wedge bonding. Measurements were performed at liquid Helium temperatures (4.2 K) using an ac voltage at 2333 Hz with 10 mV amplitude (rms) superimposed to a DC voltage (swept to align the band with the QD energy levels) and a lock-in amplifier. The QD density $\rho_{QD}$ extracted from the diode-background subtracted capacitance $C$ of the finite twofold degeneracy of the lowest energy orbital states is given by

$$\rho_{QD} = \frac{\lambda}{2e\,A_G} \int C\,dV$$

where $\lambda$ is the ratio of the distances between the n-doped back contact and i) the Schottky gate and ii) the QDs, $e$ the elementary charge, $V$ the dc gate voltage and $A_G$ the surface area of the gate[1].

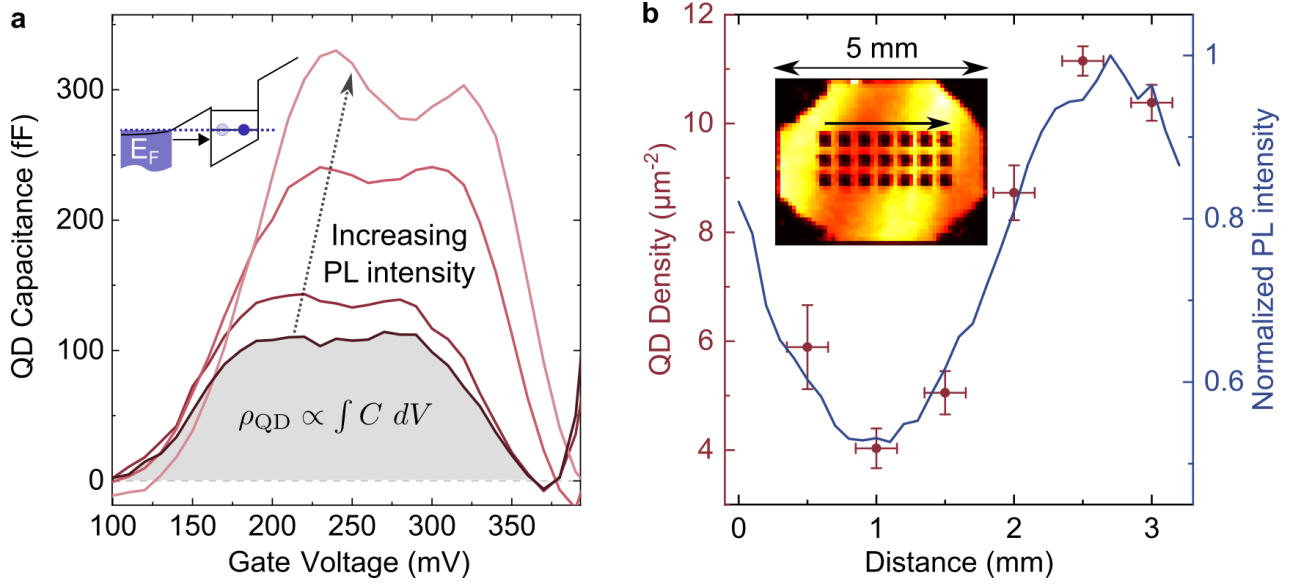

**Supp. Fig. 2 | Complementary QD density measurements using capacitance voltage spectroscopy. a**, Diode-background subtracted Capacitance-Voltage spectra of buried QDs along the PL intensity modulation. The inset shows a sketch of the sample bandstructure at a voltage, where electrons can tunnel from the Fermi level of the n-doped back contact $E_F$ to the QD ground state. **b**, QD densities deduced by C-V spectroscopy (red dots) and PL intensity (blue line) are plotted versus distance on the sample. The inset shows a PL map of the processed area (same color scheme as in Supp. Fig. 1). Black regions are metallization of the back-contacts in the corners and Schottky-gates, respectively.

**Detailed study of local contrast**

Supplementary Fig. 3a shows a PL map of the same sample as used in Fig. 1c. In Supp. Fig. 3b, we plot the Michelson-contrast of the PL map by comparing spectrally integrated ranges for high and low intensity by creating polynomial envelope functions:

$$c_{\text{Michelson}} = \frac{I_{\text{high}} - I_{\text{low}}}{I_{\text{high}} + I_{\text{low}}},$$

with $I_{\text{high}}$ the upper and $I_{\text{low}}$ the lower envelope (cf. Supp. Fig. 4c). In all samples, highest local contrast is found in regions with an InAs amount close to the critical amount $\Theta_C$, *i.e.* the onset of QD nucleation.

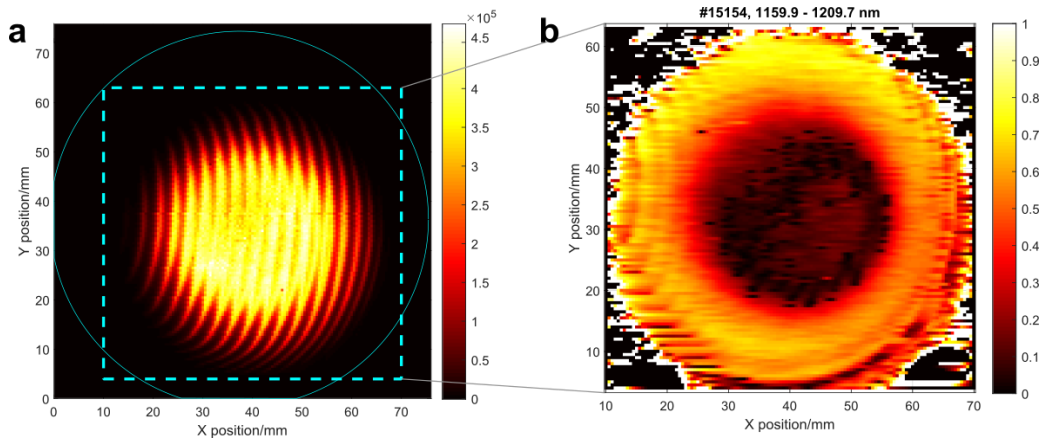

**Supplementary Fig. 3 | Photoluminescence and Michelson-contrast for $t_{\text{anneal}} = 0$ s sample. a**, PL map. **b,** Michelson contrast of the marked region in **a** (blue dashed box).

In Supp. Fig. 4 we show a sample exhibiting a high contrast. At contrast values approaching 1 (in the region of 20 – 30 mm), differentiating extremely high contrasts is difficult.

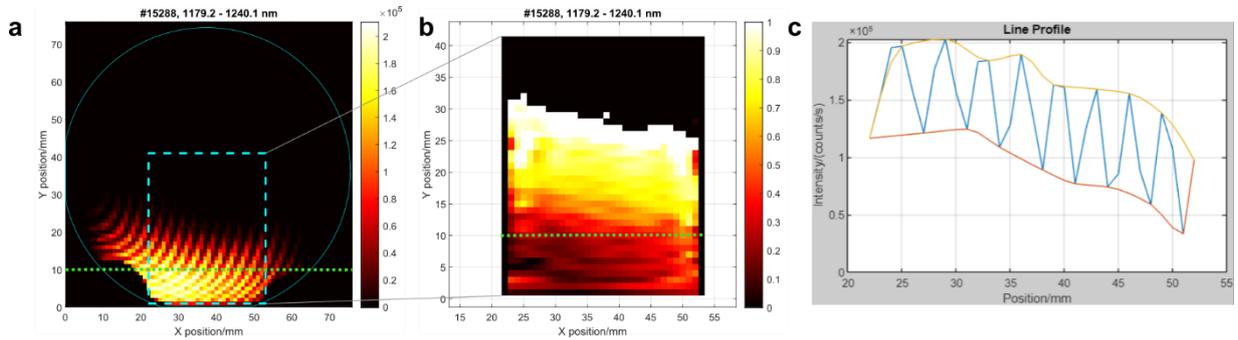

**Supp. Fig. 4 | Photoluminescence and Michelson-contrast for a high contrast sample**. **a**, Full wafer PL map. **b**, Michelson-contrast of the marked region in **a** (blue dashed box). Note that the QD nucleation onset region is located at y = 20-30 mm, where due to scaling no luminescence is visible in **a**. **c**, PL intensity (blue line) along the green dotted line from subfigure **a** and upper (yellow) and lower (red) envelope functions.

Supplementary Fig. 5 uses the Weber contrast $c_{\text{Weber}} = (I_{\text{high}} - I_{\text{low}}) / I_{\text{low}}$, since it is more suitable for visualizing high contrast values. High Weber-contrasts $c_{\text{Weber}} > 100$ are found, as plotted in Supp. Fig. 5b. Since these high contrasts are found at very low QD densities, the comparison in Fig. 1d was made from locations at similar QD densities for the different samples. As a measure for QD density, we used the relation of the s-s and p-p transitions of the QD PL emission. In our setup, locations of equal transition counts correspond to a medium density of ~ 10 QDs/µm$^2$.

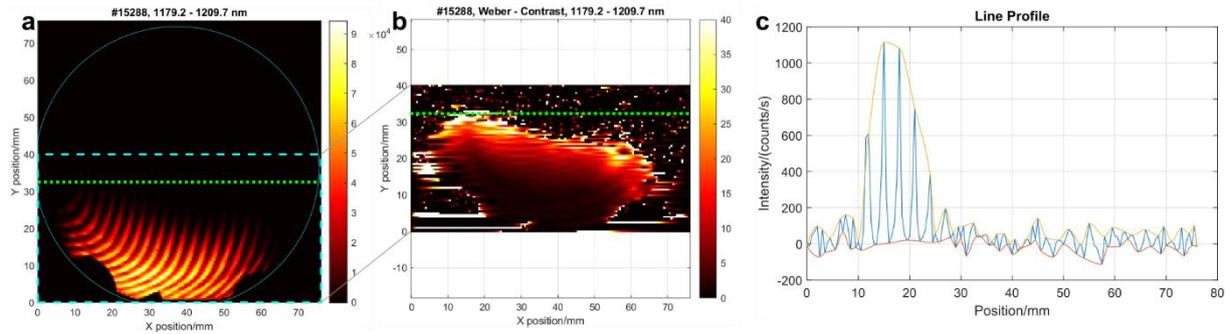

**Supp. Fig. 5 | Photoluminescence and Weber-contrast for a high contrast sample**. **a**, Full wafer PL map. **b**, Weber-contrast of the marked region in **a** (blue dashed box). Note that the QD nucleation onset region is located at y = 20-30 mm, where due to scaling no luminescence is visible in **a**. **c**, PL intensity (blue line) along the green dotted line (at the onset of QD nucleation) from subfigure **a** and upper (yellow) and lower (red) envelope functions.

**Hints of density modulation for local droplet etched quantum dots**

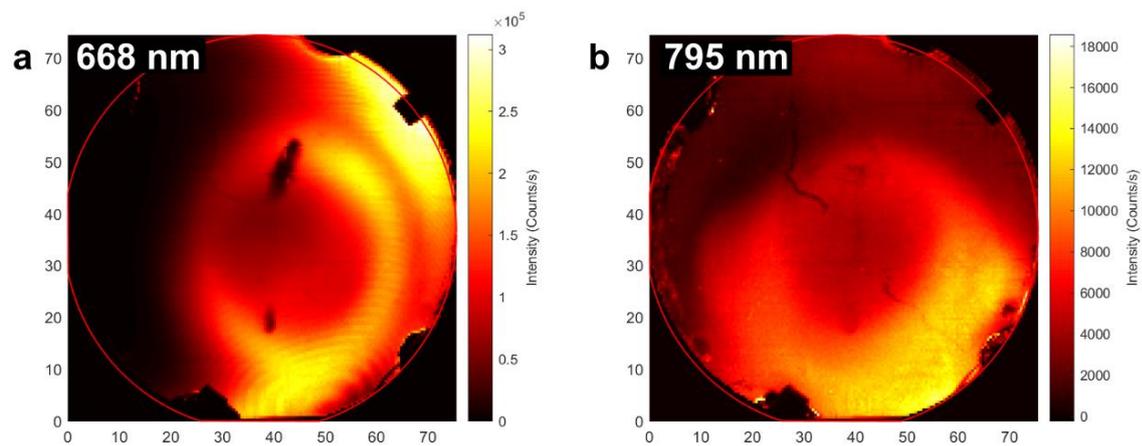

**Supplementary Fig. 6 | Photoluminescence measurements performed on local Al droplet etch QD. a**, PL map at 668 nm. **b,** PL map of the GaAs quantum dots at 795 nm in the same sample.

In Supp. Fig. 6, we present PL-data of local droplet etched QDs. We grow an AlGaAs PDL with a nominal thickness of 30 nm at the centre of the wafer oriented from bottom right to top left. Despite local droplet etching being not a Stranski-Krastanov growth method, we find a faint modulation of the intensity in the PL map. Possible reasons for the weak contrast are a high growth temperature and long annealing breaks necessary for local droplet etch dot epitaxy after the PDL growth[2].

**Wafers used for assessing and benchmarking the quality of the QDs**

Supplementary Fig. 7 presents sample wafer #14843 which consists of a high reflectivity distributed Bragg reflector (DBR) structure with a n-i-p-diode on top. After the n-layer, a tunnel barrier acting as a pattern defining layer was grown under substrate rotation. A PL map of an unprocessed wafer part is presented in Supp. Fig. 7a. Supplemantary Fig. 7b shows a simulation of the circular grown PDL. In the simulation the density modulation is based on layer thicknesses determined from quantum well emission. We attribute layers with integer monolayers thicknesses a standard nucleation probability and half monolayer thicknesses an enhanced probability with sinusoidal modulation in between. This simple assumption shows qualitative agreement with the experimental wafer map. As a comparison, in Supp. Fig. 7c, a simulation without a PDL is shown, which would result in an unmodulated distribution of QDs. In Supp. Fig. 7d we show a resonance excitation scan of a neutral exciton transition demonstrating a linewidth of 1.37 µeV which corresponds to ~1.15 times the natural linewidth, determined by lifetime measurements performed on QDs located in an area marked by the red box. By looking at a positively charged trion $X^+$ in a QD present in the sample under pulsed excitation we measure the second order autocorrelation function and demonstrate anti-bunching with a $g^2(0) = 0.01$, as can be seen in Supp. Fig. 7e. In Supp. Fig. 7f we show the raw indistinguishability measurements by comparing the central peak area at $\tau = 0$ delay for co- and cross-polarized photons and demonstrate a state-of-the-art indistinguishability with a HOM-visibility of V = 0.94. Further measurements of dots from this wafer shown by Najer et al.[3], as well as Tomm et al.[4], have demonstrated brightness in the GHz rate regime, an efficiency of more than 50% and coherence over thousands of consecutively emitted photons.

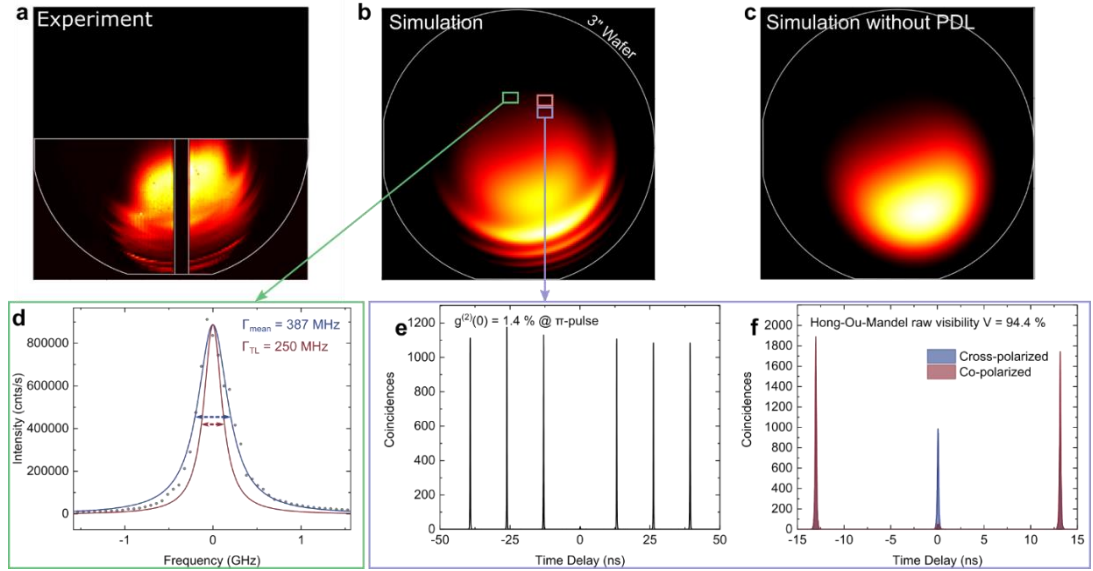

**Supp. Fig. 7 | Photoluminescence and quantum optics measurements of wafer #14843 with a circular PDL and patterned QD distribution. a,** PL map of the quantum dots. **b,** Simulation of the circular PDL and resulting quantum dot distribution. The green box marks the area for the resonant linewidth scan, the purple box marks the area for anti-bunching and indistinguishability measurements. **c,** Simulation of the same layer structure but without PDL. A homogeneous dot distribution is visible. **d,** Resonant linewidth scan of a neutral exciton. Blue dots represent the measured values, the red line is a Lorentz fit of the data. The left inset shows an SEM image of the nanophotonic waveguide. The right inset shows lifetime measurements with an exponential fit and the decay rate $\gamma_{decay}$. **e,** Second-order auto correlation measurement of a positively charged trion $X^+$ showing a $g^2(0) = 0.01$. **f,** Two-photon correlation measurement showing an HOM-visibility between co- (red area) and cross-polarized photons (blue area) of V = 0.94.

Supplementary Fig. 8 presents sample wafer #14813 which also consists of a DBR structure with a n-i-p-diode on top. After the n-layer, a tunnel barrier acting as a pattern defining layer was grown. Samples from this wafer have been processed and used to measure physical quantities, such as the single-electron tunneling rate into a quantum dot, that require an ultra-low noise environment for the dots[5-9].

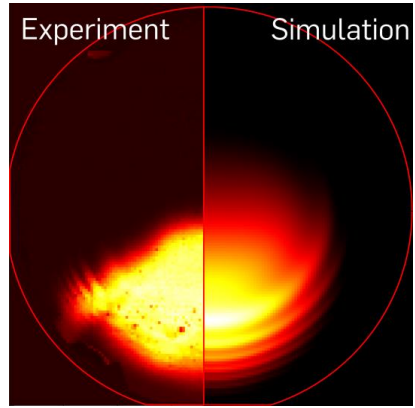

**Supp. Fig. 8 | Photoluminescence measurements and simulation of Wafer #14813.** Devices processed from areas with low PL intensity (and therefore low QD density) have been used for the ultra-low-noise measurements[5-9].

Supplementary Fig. 9 presents sample wafer #15027 which consists of QDs embedded in a n-i-p diode. A sacrificial AlGaAs layer below the n-layer allows for fabrication of a thin photonic membrane. The tunnel barrier grown after the n-doped contact layer serves as a circular PDL. In Supp. Fig. 9a we show a PL map that demonstrates the quantum dot density modulation on the wafer. In Supp. Fig. 9b we show a simulation of this area which agrees with the measurement. Supplementary Fig. 9c shows a full wafer simulation with colored boxes indicating from which area samples have been processed and measured. In Supp. Fig. 9d we demonstrate the excellent noise-free quality of the dots by showing a resonant linewidth of 568 MHz corresponding to ~1.14 times the natural linewidth. We then verify the state-of-the-art quantum optics properties, as can be seen in Supp. Fig. 9e, f. We achieve an anti-bunching of $g^{(2)}(0) = 0.02$ and a HOM-visibility of V = 0.93. Further measurements using dots from this wafer (Supp. Fig. 9c, purple box) demonstrate the integrability with excellent quantum photonic properties, such as near transform-limited linewidth and a coherent spin-photon interface by Uppu et al.[10,11], Pedersen et al.[12] and Appel et al.[13].

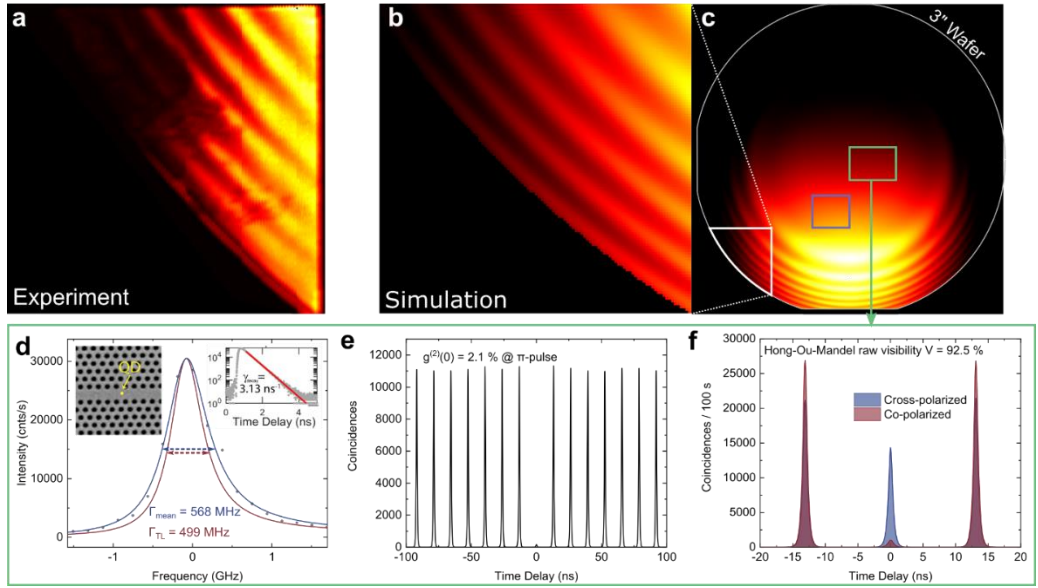

**Supp. Fig. 9 | Photoluminescence and quantum optics measurements and simulation of Wafer #15027. a,** PL map of a waferpiece. **b,** Simulation zoom-in of the layer structure for the experimentally measured wafer piece. **c,** Simulation of the full 3" wafer. The area with a white outline corresponds to the area shown in subfigure b. The green box marks the area for the quantum optics measurements and the purple box marks the area from which further measurements have been performed[10-13]. **d,** Resonant linewidth scan showing a linewidth of 568 MHz. **e,** Pulsed second-order autocorrelation showing an anti-bunching of $g^2(0) = 0.02$. **f,** Two-photon interference measurement showing a HOM-visibility of $V = 0.93$.

**Size analysis of quantum dots using atomic force microscopy**

Supplementary Figure 10 shows the width of the QDs, measured as full width at half maximum (FWHM) (red circles) and the total height of the QDs (blue circles) and the number of QDs found in each 2 x 2 µm² area as discussed in Fig. 3 in the main text. We observe slightly larger QDs at low QD densities. Between low density QD regions, the dot size seems to plateau. We find QDs tend to be slightly larger at low densities (height = (16+-0.4) nm, FWHM = (24.9+-0.1) nm) compared to higher densities (height = (15.4+-0.1) nm, FWHM = (24.6+-0.1) nm).

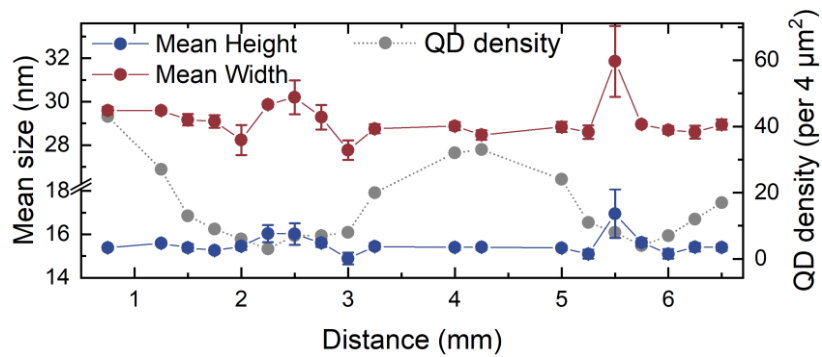

**Supp. Fig. 10 | Mean size distributions of quantum dots along a PDL.** Mean full width at half maximum (blue) and mean absolute height (red) of quantum dots along a density modulation. Error bars indicate the standard error of the mean value. The total number of QDs in a 2 x 2 µm² AFM image is shown in grey. Lines between the points are a guide to the eye.

# References (Supplementary Information)

1   Drexler, H., Leonard, D., Hansen, W., Kotthaus, J. P. & Petroff, P. M. Spectroscopy of quantum levels in charge-tunable InGaAs quantum dots. *Phys Rev Lett* **73**, 2252-2255, doi:10.1103/PhysRevLett.73.2252 (1994).
2   Zhai, L. *et al.* Low-noise GaAs quantum dots for quantum photonics. *Nature Communications* **11**, 4745, doi:10.1038/s41467-020-18625-z (2020).
3   Najer, D. *et al.* A gated quantum dot strongly coupled to an optical microcavity. *Nature* **575**, 622-627, doi:10.1038/s41586-019-1709-y (2019).
4   Tomm, N. *et al.* A bright and fast source of coherent single photons. *Nature Nanotechnology*, 1-5 (2021).
5   Lochner, P. *et al.* Contrast of 83% in reflection measurements on a single quantum dot. *Scientific Reports* **9**, 8817, doi:10.1038/s41598-019-45259-z (2019).
6   Kurzmann, A., Ludwig, A., Wieck, A. D., Lorke, A. & Geller, M. Auger Recombination in Self-Assembled Quantum Dots: Quenching and Broadening of the Charged Exciton Transition. *Nano Letters* **16**, 3367-3372, doi:10.1021/acs.nanolett.6b01082 (2016).
7   Kurzmann, A., Ludwig, A., Wieck, A. D., Lorke, A. & Geller, M. Photoelectron generation and capture in the resonance fluorescence of a quantum dot. *Applied Physics Letters* **108**, 263108 (2016).
8   Kurzmann, A. *et al.* Optical detection of single-electron tunneling into a semiconductor quantum dot. *Physical Review Letters* **122**, 247403 (2019).
9   Kurzmann, A. *et al.* Optical blocking of electron tunneling into a single self-assembled quantum dot. *Physical review letters* **117**, 017401 (2016).
10  Uppu, R. *et al.* On-chip deterministic operation of quantum dots in dual-mode waveguides for a plug-and-play single-photon source. *Nature Communications* **11**, 3782, doi:10.1038/s41467-020-17603-9 (2020).

11      Uppu, R. *et al.* Scalable integrated single-photon source. *Science Advances* **6**, eabc8268 (2020).
12      Pedersen, F. T. *et al.* Near Transform-Limited Quantum Dot Linewidths in a Broadband Photonic Crystal Waveguide. *ACS Photonics* **7**, 2343-2349, doi:10.1021/acsphotonics.0c00758 (2020).
13      Appel, M. H. *et al.* Coherent Spin-Photon Interface with Waveguide Induced Cycling Transitions. *Physical Review Letters* **126**, 013602 (2021).